\documentclass[aps,preprint,showpacs]{revtex4}
\usepackage{amsfonts}
\usepackage{amsmath}
\usepackage{amssymb}

\usepackage{graphicx}
\usepackage{epsfig}

\newcommand{\bfr}{{\mbox{\boldmath$\rho$}}}

\begin{document}

\title{On the dependence of one-photon--two-electron differential cross sections on recoil momenta}
\author{M. Ya. Amusia $^{1,2}$, E. G. Drukarev $^{1,3}$, E. Z. Liverts $^1$, A. I. Mikhailov $^3$}
\affiliation{$^1$The Racah Institute of Physics,
The Hebrew University of Jerusalem, Jerusalem 91904 Israel\\
$^2$ A. F.Ioffe Physical-Technical Institute, St. Petersburg 194021 Russia\\
$^3$ B. P. Konstantinov Petersburg Nuclear Physics Institute, Gatchina,
St. Petersburg, 188300 Russia}

\begin{abstract}

We calculate the distributions in recoil
momenta and their energy distributions
for the high energy non-relativistic double photoionization of helium
caused by the quasifree mechanism of the process. The distributions
obtain local maxima at small values of the recoil momenta. This is in
agreement with the earlier predictions and with recently obtained experimental
data. We obtained also the angular correlations, which reach the largest value
in the ``back-to-back" configuration of the photoelectrons.
Our analysis is valid in all high energy nonrelativistic region. Particular
equations are true for the case when the wavelength of the photon exceeds
strongly the size of the atom. We present numerical results for
the photon energies in the region of 1\,keV, employed in the recent experiments.

\end{abstract}

\date{\today}
\maketitle

\section{Introduction}

The recent measurement of the yield of double charged ions in photoionization of helium \cite{7a}-\cite{7}
confirmed the exitance of the
\textit{quasifree mechanism} (QFM) of the double photoionization
which was predicted in \cite{1}.
The differential cross sections of the double photoionization were calculated earlier in a number of papers
\cite{1}--\cite{6}, where the authors studied the distributions in
characteristics of each photoelectron.  In the pioneering experiments
\cite{7a}-\cite{7} the distribution in momentum transferred to the
nucleus $ q$ (recoil momentum) was measured.  Thus the problem of
calculation of such distributions as  $d\sigma^{2+}/dq^2d\varepsilon$
with $\varepsilon$ the energy of one of the photoelectrons and  $d\sigma^{2+}/dq^2$ became actual.

In the present paper we calculate these differential cross sections for
the double photoionization of helium at high values of the photon energies,
corresponding, however, to nonrelativistic energies of the photoelectrons.
We present the results for the differential cross sections $d\sigma^{2+}/dq^2d\varepsilon$
We trace the dependence of these characteristics on the photon energy $\omega$.

Recall that the experiments \cite{7a}-\cite{7} in which the photons
carried the energies $450$\,eV, $800$ eV and $900$\,eV demonstrated that
the distribution of outgoing electrons obtains a surplus at small
$q$ of about 2 a.u. The kinematics of these experiments enables to
separate the non-dipole contributions at small values of $q$. Thus
the observed surplus is entirely due to the non-dipole terms.

By that time only two mechanisms of the process were known. In both
of them the electron which interacted with the photon directly
obtained almost all the incoming photon energy $\omega$. In the
first one, known as the \textit{shake-off} the secondary electron is pushed
to continuum by to the sudden change of the effective field. In
the second, called the \textit{knock-out} mechanism, the photoelectron
inelastically collides with the bound one, sharing the photon energy.
Both mechanisms contain the single photoionization as the first step.
This process can not take place on a free electron.
Thus the momentum $q$ which is transferred in this step to the nucleus
exceeds strongly the averaged momentum of the bound electron $\mu$,
in the case of the high energy photon
\begin{equation}
\label{1}
\omega \gg I,
\end{equation}
($I$ is the single-particle ionization potential). Of course, there is a configuration
in which the second electron transfers momentum $q_1 \gg \mu$ such as
$|{\bf q}+{\bf q_1}| \sim \mu$. However since each act of transferring a large momentum
$q \gg \mu$ leads to an additional small factor \cite{8}, \cite{9} its probability is very small.
The distribution $d\sigma^{2+}/dq^2d\varepsilon$ provided by these two mechanisms peaks at
$q \approx (2m\omega)^{1/2} \gg \mu$ (m is the electron mass), becoming very small at $q \sim \mu$.
This remains true beyond the dipole approximation.

In contrast to a single electron case,
the two electrons can absorb a photon without participation
of the nucleus. In the free process ${\bf q}=0$. In the QFM small momentum $q \sim \mu$ is transferred to the nucleus,
i.e. $q$ is much smaller than the momenta of the outgoing electrons.
The distributions $d\sigma^{2+}/dq^2d\varepsilon$ and $d\sigma^{2+}/dq^2$
the have local maxima at small $q$ of the order of $\eta$. That's what was detected in \cite{7a}-\cite{7}.

Momentum ${\bf q}$ transferred to the nucleus can be written as
\begin{equation}
\label{3}
{\bf q}={\bf k}-{\bf p_1}- {\bf p_2},
\end{equation}
where ${\bf p}_{1,2}$ are momenta of the outgoing electrons, while $k$ is that of the photon. The recoil momentum $q$
can become small only if the large momenta of the outgoing photoelectron with $p_i=|{\bf p}_i| \ll \mu$ compensate each other to large extent ($k=|{\bf k}|$ is always much smaller than $p_i$ while we consider the photon energies, corresponding to nonrelativistic photoelectrons). Hence the values of $p_i$ should be close, i.e. $p_1 \approx p_2 \approx \sqrt {mE} $ with $E$ the sum of the energies of the photoelectrons. Thus in QFM the bound electrons exchange by  small momentum $q \sim \mu$ with the nucleus and by large momentum of the order $p_i \gg \mu$ between themselves.

We calculate the amplitude of the QFM in the lowest order of expansion in powers of $q/p_i$. This corresponds to expansion of the bound state wave function in the lowest order in powers of $r_{12}/r_i$, with $r_i$ standing for the distance between the electron and the nucleus, while $r_{12}$ is the interelectron distance.
We consider the high energy photons, corresponding however to nonrelativistic energies of the outgoing electrons. Thus we assume that $\omega \ll m$. Having in mind future extension of the analysis to the relativistic case we employ the relativistic units $\hbar =c=1$.

Since $q \sim \mu$ the higher terms of expansion in powers of $q/p_i$ are of the same order as those coming from the interactions between the photoelectrons and the nucleus. However, they are of quite different physical origin. Thus we include interactions of the nucleus with the electrons exactly, describing the latter by the nonrelativistic functions of the Coulomb field. Interaction between the outgoing electrons is  proportional to the square of its Sommerfeld parameter
$\xi_{ee}=\alpha/v$, where $v$ is their relative velocity.
For the energies of the order $1$ keV, employed in the experiments \cite{7a}-\cite{7} this interaction
provides a correction of the order of 2 $\%$ and can be neglected.

Direct relation of the QFM to the behavior of the bound state wave function
$\Psi(r_1,r_2, r_{12})$ at small distance $r_{12}$ was demonstrated in \cite{8}. It was shown that the QFM amplitude  contains the factor $\partial
\Psi/\partial r_{12}$ at $r_{12}=0$, which is connected to the
function $\Psi(r_1,r_1, r_{12}=0)$ by the Kato cusp condition \cite{9}. The latter appears to be very important
for calculation of the QFM amplitude \cite{4}.
We employ very precise wave function \cite{10} which satisfy also the Kato cusp conditions. We use an analytical function which approximate these wave functions at the electron coalescence line $r_{12}=0$ very accurately \cite{11}.

We include only the quadrupole part of the electron-photon interaction.
Note that the amplitude contains also the dipole terms proportional to the product $({\bf e} \cdot {\bf q})$
with ${\bf e}$ the vector of the photon polarization. However,
at least in the leading approximation it is canceled by the contribution in which the electrons exchange by large momentum in the final state \cite{4}. Anyway, in the experiments \cite{7a}-\cite{7} the observations were carried out in the
plane where $({\bf e} \cdot {\bf q})=0$. Thus we can focus on the quadrupole contribution.

Besides the conservation of the linear momentum expressed by Eq.(\ref{3}) we write the energy conservation condition
\begin{equation}
\label{2}
\omega-I=E; \quad E=\varepsilon_1+\varepsilon_2,
\end{equation}
where $\varepsilon_{i}=p_i^2/2m$ ($i=1,2$) are the photoelectron energies.
Note that in our system of units $\omega=|{\bf k}|$.

To simplify the calculations we restrict ourselves to the case when the photon wave length is much larger than the size of the bound state, i.e.
\begin{equation}
\label{4}
\omega \ll \mu.
\end{equation}
For the atom of helium this means that $\omega \ll 6$ keV.
Under this condition Eq.(\ref{3}) can be written as
 \begin{equation}
{\bf q}=-{\bf p_1}- {\bf p_2},
\label{5}
\end{equation}
in the lowest order of expansion in powers of $k$.

Momentum $q$ can become as small as $\mu$ only if momenta of the outgoing electrons almost compensate each other,
i.e. $|{\bf p_1}+{\bf p_2}| \sim \mu \ll p_{1,2}$. Hence, the photoelectrons are emitted mostly "back-to-back, with $t \equiv ({\bf
p_1} \cdot {\bf p_2})/p_1p_2$ close to $-1$, while the values $p_1 \approx p_2$, i.e. $|p_1-p_2| \ll p_{1,2}$.
Thus the relative difference of the energies of the outgoing electrons
\begin{equation}
\beta \equiv \frac{|\varepsilon_1-\varepsilon_2|}{E}
\label{5a}
\end{equation}
should be small.
Since $ q \geq |p_1-p_2|$ we find
\begin{equation}
\beta <\frac{q}{(mE)^{1/2}} \ll 1.
\label{6}
\end{equation}
This equation is presented in the lowest order in $\beta$.

Besides the distributions $d\sigma^{2+}/dq^2d\varepsilon$
and  $d\sigma^{2+}/dq^2$ we calculate the differential cross sections $d\sigma^{2+}/dtd\varepsilon$
and  $d\sigma^{2+}/dt$. We present the numerical data for the photons carrying the energy of about $1$ keV, employed in the experiments \cite{7a}-\cite{7}.

Note that this approach was used in \cite{12} for calculation of the distributions
$d\sigma^{2+}/dtd\varepsilon$ and
$d\sigma^{2+}/d\varepsilon$
at the point of the peak $t=-1$.
In other words in \cite{12} the height of the peak of this
distributions was found. In the present paper we calculate the shape of the peaks.

\section{General equations}

The differential cross section of the double photoionization can be written as
\begin{equation}
d\sigma^{2+}=\frac{1}{2\omega}|F({\bf k}, {\bf p_1},{\bf p_2})|^2d\Gamma.
\label{7}
\end{equation}
Here $F({\bf k}, {\bf p_1},{\bf p_2})$ is the amplitude of the process. Averaging over polarizations of the photon is assumed. The last factor is the phase volume
\begin{equation}
d\Gamma=2\pi\delta(\omega-I-\varepsilon_1-\varepsilon_2)\frac{d^3p_1}{(2\pi)^3}\frac{d^3p_2}{(2\pi)^3}.
\label{8}
\end{equation}
Employing Eq.(\ref{5}) we can present
\begin{equation}
d\Gamma=\delta(\omega-I-2\varepsilon_1-\frac{p_1q_z}{m}-\frac{q^2}{2m})\frac{dq^2dq_z}{4\pi}\frac{d^3p_1}{(2\pi)^3}.
\label{9}
\end{equation}
with $z$ the direction of momentum ${\bf p_1}$.
Using $\delta$-function for integration over $q_z$ we can write
\begin{equation}
d\sigma^{2+}=|F({\bf k}, {\bf p_1},{\bf p_2})|^2\frac{\theta(q/p-\beta)}{8\pi}\frac{Em^2}{\omega}
\frac{d\beta}{2\pi}\frac{d\Omega}{4\pi}\frac{dq^2}{2\pi},
\label{10}
\end{equation}
with $\Omega$ the solid angle of the photoelectron with momentum ${\bf p_1}$, $p=(mE)^{1/2}$.

The amplitude of the process can be written as
\begin{equation}
F({\bf k}, {\bf p_1},{\bf p_2})=\langle\Psi_f(1,2)|\gamma_1+\gamma_2|\Psi_i(1,2)\rangle,
\label{11}
\end{equation}
with the numbers $1$ and $2$ denote the variables corresponding to two electrons, $\Psi_{i,f}$ are the wave functions of the initial and final states,
\begin{equation}
\gamma=(4\pi\alpha)^{1/2}e^{i({\bf k} \cdot {\bf r})}\frac{-i({\bf e} \cdot {\bf \nabla})}{m},
\label{12}
\end{equation}
where ${\bf e}$ is the vector of polarization of the photon, $({\bf e} \cdot {\bf k})=0$.
Recall that we shall pick only the quadrupole terms of interaction between the photon end electron.
For further evaluation we denote
\begin{equation}
F({\bf k}, {\bf p_1},{\bf p_2})=(4\pi\alpha)^{1/2}M({\bf k}, {\bf p_1},{\bf p_2}).
\label{13}
\end{equation}
As we said earlier, we describe the final state by the function
\begin{equation}
\Psi_f({\bf r_1}, {\bf r_2})=\frac{1}{\sqrt 2}\Big(\psi_{1}({\bf r_1})\psi_{2}({\bf r_2})+\psi_{2}({\bf r_1})\psi_{1}({\bf r_2})\Big),
\label{14}
\end{equation}
where $\psi_i$ are the single-particle nonrelativistic Coulomb field function with asymptotic momenta ${\bf p_i}$.
We shall need the functions
\begin{equation}
\psi^*_{\bf p_i}({\bf r})=e^{-i({\bf p_i}\cdot {\bf r})}X({\bf p_i},\xi_i,{\bf r}); \quad i=1,2
\label{15}
\end{equation}
Here
\begin{equation}
X({\bf p_i},\xi_i,{\bf r})=N(\xi_i)_1F_1(i\xi_i,1,ip_ir+i({\bf p_i}\cdot{\bf r})),
\label{16}
\end{equation}
while
\begin{equation}
N(\xi_i)=\Big(\frac{2\pi\xi_i}{1-e^{-2\pi\xi_i}}\Big)^{1/2},
\label{17}
\end{equation}
with
\begin{equation}
\xi_i=\frac{\eta}{p_i}; \quad \eta=m\alpha Z.
\label{18}
\end{equation}
Here $Z$ is the charge of the nucleus. Note that in the hydrogenlike approximation $\eta$ is the averaged momentum of the electron in the $1s$ state.

Thus
\begin{equation}
M({\bf k}, {\bf p_1},{\bf p_2})=\sqrt{2}\Big(A({\bf k}, {\bf p_1},{\bf p_2})+({\bf p_1} \leftarrow \rightarrow  {\bf p_2})\Big);
\label{19}
\end{equation}
$$A({\bf k}, {\bf p_1},{\bf p_2})=
\int d^3r_1d^3r_2e^{i({\bf k}\cdot {\bf r_1})-i({\bf p_1}\cdot {\bf r_1})
-i({\bf p_2}\cdot {\bf r_2})}X_1({\bf r_1})X_2({\bf r_2})({\bf e} \cdot {\bf \nabla_{r_1}})\Psi_i({\bf r_1}, {\bf r_2})\equiv A.$$
Here we denoted
\begin{equation}
X_i({\bf r})=X({\bf p_i},\xi_i,{\bf r}).
\label{19a}
\end{equation}
Introduce
\begin{equation}
{\bf R}=\frac{{\bf r}_1+{\bf r}_2}{2};\quad {\bfr}={\bf r}_2-{\bf r}_1
\label{20}
\end{equation}
Presenting
\begin{equation}
{\bf r}_1={\bf R}-{\bfr}/2;\quad {\bf r_2}={\bf R}+{\bfr}/2; \quad {\bf \nabla}_{r_1}=\frac{1}{2}{\bf \nabla}_{R}-{\bf \nabla}_{\rho},
\label{21}
\end{equation}
and
\begin{equation}
\Psi_i({\bf r_1}, {\bf r_2})=\Psi({\bf R}, {\bfr})
\label{22}
\end{equation}
we obtain
\begin{equation}
A=A_1+A_2,
\label{23}
\end{equation}
with
\begin{equation}
A_1=\frac{i}{m}\int d^3Rd^3\rho e^{-i({\bf a} \cdot {\bfr})+i({\bf q} \cdot {\bf R})}X_1({\bf R}-\frac{\bfr}{2})
X_2({\bf R}+\frac{\bfr}{2})({\bf e}\cdot {\bf \nabla}_{\rho})\Psi({\bf R}, {\bfr}),
\label{24}
\end{equation}
while
\begin{equation}
A_2=-\frac{i}{2m}\int d^3Rd^3\rho e^{-i({\bf a} \cdot {\bfr})+i({\bf q} \cdot {\bf R})}X_1({\bf R}-\frac{\bfr}{2})
X_2({\bf R}+\frac{\bfr}{2})({\bf e }\cdot {\bf \nabla}_{R})\Psi({\bf R}, {\bfr}),
\label{25}
\end{equation}
where
\begin{equation}
{\bf a}=\frac{{\bf p}_1-{\bf p}_2+{\bf k}}{2}; \quad a=|{\bf a}|
\label{26}
\end{equation}

Since $a \gg q $, the integrals on the right hand sides of Eqs.(\ref{24}), (\ref{25}) are saturated by
$R \sim 1/q \gg \rho$. Thus we can put ${\bfr}=0$ in the functions $X_i$. This provides
\begin{equation}
A_1=\frac{({\bf e }\cdot {\bf a})}{m}\int d^3Re^{i({\bf q} \cdot {\bf R})}X_1({\bf R})
X_2({\bf R})\int d^3\rho e^{-i({\bf a} \cdot {\bfr})}\Psi({\bf R}, {\bfr}),
\label{26a}
\end{equation}
Now we expand the wave function
\begin{equation}
\Psi({\bf R}, {\bfr})=\Psi(R, \tau, \rho)=\Psi(R, 0,0 )+\tau\Psi'_{\tau}(R, \tau, 0)+\rho\Psi'_{\rho}(R, 0, \rho)+0(\rho^2).
\label{27}
\end{equation}
Here $\tau=({\bf R}\cdot{\bfr})$, the derivatives are taken at $\tau=\rho=0$.

We can calculate the integral over $\rho$ multiplying the integrand by $e^{-\nu\rho}$ and putting $\nu=0$ in the final step.
\begin{equation}
\int d^3\rho e^{-i({\bf a} \cdot {\bfr})}\Psi({\bf R}, {\bfr})=
\int d^3\rho e^{-i({\bf a} \cdot {\bfr})-\nu\rho}\Psi({\bf R}, {\bfr}=0)=\int d^3\rho e^{-i({\bf a} \cdot {\bfr})-\nu\rho}\rho\Psi_{\rho}'(R, 0, \rho=0)=
\label{27a}
\end{equation}
$$-\frac{8\pi\Psi_{\rho}'(R, 0, \rho=0)}{a^4}$$.

The derivative $\Psi_{\rho}$ at $\rho=0$ is related to the wave function by the Kato cusp condition \cite{11}

\begin{equation}
lim_{\rho \rightarrow 0}r_0\Psi_{\rho}'({\bf R }, {\bfr})=\frac{1}{2}\Psi(R,0),
\label{27b}
\end{equation}
where $r_0=1/m\alpha$ is the Bohr radius. It is identical to similar relation for the wave function presented in variables $r_1$,$r_2$,$\rho$. Introducing
\begin{equation}
\Phi(R)=\Psi({\bf R}, {\bfr}=0),
\label{27c}
\end{equation}
we can write
\begin{equation}
A_1=\frac{4\pi\alpha}{a^4}({\bf e }\cdot {\bf a})S_1(q),
\label{28}
\end{equation}
with
\begin{equation}
S_1(q)=\int d^3Re^{i({\bf q} \cdot {\bf R})}X_1({\bf R})
X_2({\bf R})\Phi(R),
\label{28a}
\end{equation}
with the functions $X_i({\bf R})$ defined by Eq.(\ref{19a}).

Combining Eqs. (\ref{13}, \ref{19}, \ref{23}, \ref{26}) we find for the quadrupole terms of the amplitude
\begin{equation}
F({\bf k}, {\bf p_1},{\bf p_2})=(4\pi\alpha)^{3/2}4\sqrt{2}\frac{({\bf e} \cdot {\bf n})({\bf k} \cdot {\bf n})}{p^4}S_1(q).
\label{28b}
\end{equation}
After averaging over the photon polarization and integration over the angles Eq.(\ref{10})
takes the form
\begin{equation}
\label{50}
\frac{d^2 \sigma}{dq^2d\beta}=\frac{2^7}{15}\alpha^3\frac{\omega}{
E^4}|S_1(q)|^2.
\end{equation}

In order to calculate $S_1(q)$ we employ the presentation of the function
\begin{equation}
\Phi(R)=c_1e^{-\lambda_1 R}+c_2e^{-\lambda_2 R},
\label{27h}
\end{equation}
with numerical values of the parameters
$$c_1=0.380\zeta^3, \quad c_2=0.990\zeta^3, \quad \lambda_1=5.54\zeta, \quad \lambda_2=3.41\zeta,\quad \zeta=m\alpha$$ found in\cite{11}.

The further calculations are described in Appendix. We obtain for the function $S_1(q)$ defined by Eq.(\ref{28a})
\begin{equation}
|S_1(q)|^2=|\sum_i c_i I(\lambda_i)|^2 ;\quad i=1,2.
\label{27d}
\end{equation}
with $I(\lambda_i)$ defined by Eq.(\ref{39}).

\section{Results}

Now we present the results of computations. The cross section
${d^2 \sigma}/{dq^2d\beta}$ determined by Eq.(\ref{50}) is presented in a three-dimensional Fig.1
for $\omega=800$ eV. As expected, it obtains the largest values at small $\beta \ll 1$ and in the region of small $q \sim 1 a.u.$ in agreement with the experimental results \cite{7}. This distribution at $\beta=0$, corresponding to the center of the spectrum is shown in Fig.2
for $\omega=800 eV$ and $\omega=1 $ keV. Since the effects of finite $\beta$ manifest themselves in the terms of the order $\beta^2$, there is no noticeable difference from similar figures for $\beta \neq 0$ in the QFM region due to Eq.(\ref{6}).

It is instructive also to view the energy distribution of the angular correlation
\begin{equation}
\label{51}
\frac{d^2 \sigma}{dt d\beta}=2p_1p_2\frac{d^2 \sigma}{dq^2d\beta}; \quad t=\frac{({\bf p_1} \cdot {\bf p_2})}{p_1p_2}.
\end{equation}
It is shown for $\omega= 800$  eV in the three-dimensional Fig.4. As expected,
the largest values are reached at $\beta \ll 1$ and $t$ close to $-1$, corresponding to the electrons ejected in the opposite directions ("back-to back"). For $\beta=0$ this differential cross section is shown in Fig.4
for $\omega=800 eV$ and $\omega=1 $ keV.

We calculate also the distribution in recoil momentum
\begin{equation}
\label{52}
 \frac{d \sigma}{dq^2}=\frac{1}{2}\int _0^{q/p}d\beta\frac{d^2\sigma}{dq^2d\beta},
\end{equation}
and the angular correlation
\begin{equation}
\label{53}
 \frac{d \sigma}{d t}=\frac{1}{2}\int _0^{1}d\beta\frac{d^2\sigma}{dtd\beta}.
\end{equation}
They are presented in Fig.5 and Fig.6 correspondingly. As expected, the distribution $d\sigma/dq^2$ has a local maximum
at $q$ about $1$ a.u. At $q=0$ this distribution turns to zero just because the interval of integration over $\beta$ vanishes. The angular correlation $d\sigma/dt$ has a sharp peak at $t=-1$, in agreement with the previous analysis.

\section{Summary}

We calculated the distributions in recoil momenta $q$ and their energy distribution for the high energy nonrelativistic double photoionization of helium caused by the quasifree mechanism (QFM)\cite{1}. They are closely related to the distributions in the angle between momenta of the outgoing electrons (angular correlations). As expected, the distributions in recoil momenta obtain local maxima at small $q$ of the order $1-2$ a.u., in agreement with the
results of the pioneering experiments \cite{7a}-\cite{7}. Unfortunately, the way of presentation of the results in \cite{7a}-\cite{7} does not permit to compare the quantitative results. The corresponding angular distributions obtain maxima when the photoelectrons move in the opposite directions ("back-to-back" scattering). The qualitative picture is the same for heavier atoms.

The QFM is caused by the initial state interactions and, contrary to a misleading statement in \cite{7},  is contained in the standard  Feynman  diagrams for the amplitude \cite{1}. Since the QFM is at work at small separation between the bound electrons $r_{12}$,
we described the initial state by a very precise wave function \cite{10}, employing its analytical
approximation at small values of $r_{12}$ \cite{11}. We neglected the electron interactions in the final state. The numerical results for the photon energies
in the keV region are shown in Figs.1-6. This energy region attracts attention nowadays in connection with the laser experiments. Also, the experiments \cite{7a}-\cite{7} where carried out at these energies. The approach can be applied for the double photoionization of heavier atoms.

One can obtain more precise results by direct employing of precise wave functions, i.e. those found in
\cite{5}. However, such approaches do not allow to analyze  the mechanisms of the process. On the other hand
very "accurate" wave functions (i.e. those which reproduce the value of the binding energy very accurately)
may have a wrong behavior at $r_{12} \rightarrow 0$. It was demonstrated in \cite{4}, \cite{14} that a number of
publications on the subject employing such functions contain erroneous results. More examples are given in \cite{15}.
That is why we consider our results as a necessary step in investigation of the process.

There is a number of possibilities to carry out the experimental investigation of other phenomena connected with the QFM. Outside the plane $({\bf e}\cdot {\bf q})=0$ interference between the dipole and quadrupole terms should manifest itself in the angular distributions. Also, it would be interesting to trace the $\omega$ dependence of the shape of the energy distribution. Its theoretical analysis was presented in \cite{15}.

The work was supported by the MNTI-RFBR grant 11-02-92484. One of us (EGD)
thanks for
hospitality during the visit to the Hebrew University of Jerusalem.
\def\thesection{Appendix \Alph{section}}
\def\theequation{\Alph{section}.\arabic{equation}}
\setcounter{section}{0}

\section{}
\setcounter{equation}{0}

Thus we calculate the integral
\begin{equation}
S_1(q)=\int d^3Re^{i({\bf q} \cdot {\bf R})}X_1({\bf R})
X_2({\bf R})e^{-\lambda R}=N(\xi_1)N(\xi_2)I(\lambda),
\label{30}
\end{equation}
where
\begin{equation}
I(\lambda)=\frac{-\partial J(\lambda)}{\partial \lambda}; \quad J(\lambda)=\int d^3Re^{i({\bf q} \cdot {\bf R})}F_1({\bf R})
F_2({\bf R})\frac{e^{-\lambda R}}{R},
\label{31}
\end{equation}
with
\begin{equation}
F_i=~_1F_{1}(i\xi_i,1,ip_iR+i({\bf p_i}\cdot{\bf R})),
\label{32}
\end{equation}
The integral $J(\lambda)$ was calculated in \cite{13} as
\begin{equation}
J(\lambda)=\frac{2\pi e^{-\pi \xi_1}}{\alpha_c}\Big(\frac{\alpha_c}{\gamma_c}\Big)^{i\xi_1}\Big(\frac{\gamma_c+\delta_c}{\gamma_c}\Big)^{-i\xi_2}
~_2F_1(1-i\xi_1, i\xi_2, 1,g),
\label{33}
\end{equation}
with
\begin{equation}
\alpha_c=\frac{q^2+\lambda^2}{2}; \quad \beta_c=({\bf p_2}\cdot{\bf q})-i\lambda p_2; \quad \gamma_c=
-({\bf p_1}\cdot{\bf q})+i\lambda p_1-\alpha_c; \quad \delta_c=p_1p_2-({\bf p_1}\cdot{\bf p_2})-\beta_c,
\label{34a}
\end{equation}
$$g=\frac{\alpha_c\delta_c -\beta_c\gamma_c}{\alpha_c(\gamma_c+\delta_c)}.$$
We write Eq(\ref{33}) in a more symmetric form
\begin{equation}
J(\lambda)=\frac{2\pi e^{-\pi \xi_1}}{\alpha_c}\Big(\frac{\alpha_c}{\gamma_c}\Big)^{i\xi_1}\Big(\frac{\alpha_c}{\alpha_c+\beta_c}\Big)^{i\xi_2}
~_2F_1(i\xi_1, i\xi_2, 1,h),
\label{32a}
\end{equation}
with
\begin{equation}
h=\frac{\beta_c\gamma_c-\alpha_c\delta_c}{\gamma_c(\alpha_c+\beta_c)}.
\label{35}
\end{equation}
Thus
\begin{equation}
J(\lambda)=4\pi\Lambda(\lambda)_2F_1(i\xi_1, i\xi_2, 1,h(\lambda)),
\label{36}
\end{equation}
with
\begin{equation}
\Lambda(\lambda)=\Big(q^2+\lambda^2\Big)^{-1+i\xi_1+\xi_2}\Big(p_1+p_2+i\lambda\Big)^{-i\xi_1-i\xi_2}
\Big(p_2-p_1-i\lambda\Big)^{-i\xi_1}
\Big(p_1-p_2-i\lambda\Big)^{-i\xi_2}.
\label{37}
\end{equation}
Employing
\begin{equation}
\frac{\partial}{\partial h}~_2F_1(i\xi_1, i\xi_2, 1,h)=-\xi_1\xi_2~_2F_1(i\xi_1+1, i\xi_2+1, 2,h),
\label{38}
\end{equation}
we find
\begin{equation}
I(\lambda)=\frac{8\pi\lambda}{(q^2+\lambda^2)^2}\Theta^{i(\xi_1+\xi_2)}(\lambda)T(\lambda)e^{-\pi/2(\xi_1+\xi_2)}.
\label{39}
\end{equation}
Here
\begin{equation}
\Theta(\lambda)=\frac{q^2+\lambda^2}{s(\lambda)u(\lambda)}; \quad s(\lambda)=\sqrt{(p_1+p_2)^2+\lambda^2},
\quad u(\lambda)=\sqrt{(p_1-p_2)^2+\lambda^2}
\label{40}
\end{equation}
while
\begin{equation}
T(\lambda)=\big(1-\frac{i(\xi_1+\xi_2)}{2}(1+h(\lambda))_2F_1(i\xi_1, i\xi_2, 1,h(\lambda))-
\label{41}
\end{equation}
$$-\xi_1\xi_2h(\lambda)(1-h(\lambda))_2F_1(i\xi_1+1, i\xi_2+1, 2,h(\lambda)),$$
with
\begin{equation}
h(\lambda)=1-\frac{q^2+\lambda^2}{u^2(\lambda)}
\label{42}
\end{equation}

\newpage

\begin{figure}
\begin{center}
\caption{Distribution  $d\sigma^{2+}/dq^2d\beta$ in $10^{-10}r_0^4$, $r_0=1/m\alpha$
for $\omega=800$\,eV. The recoil momentum $q$ is in atomic
units.}
\epsfxsize=20cm\epsfbox{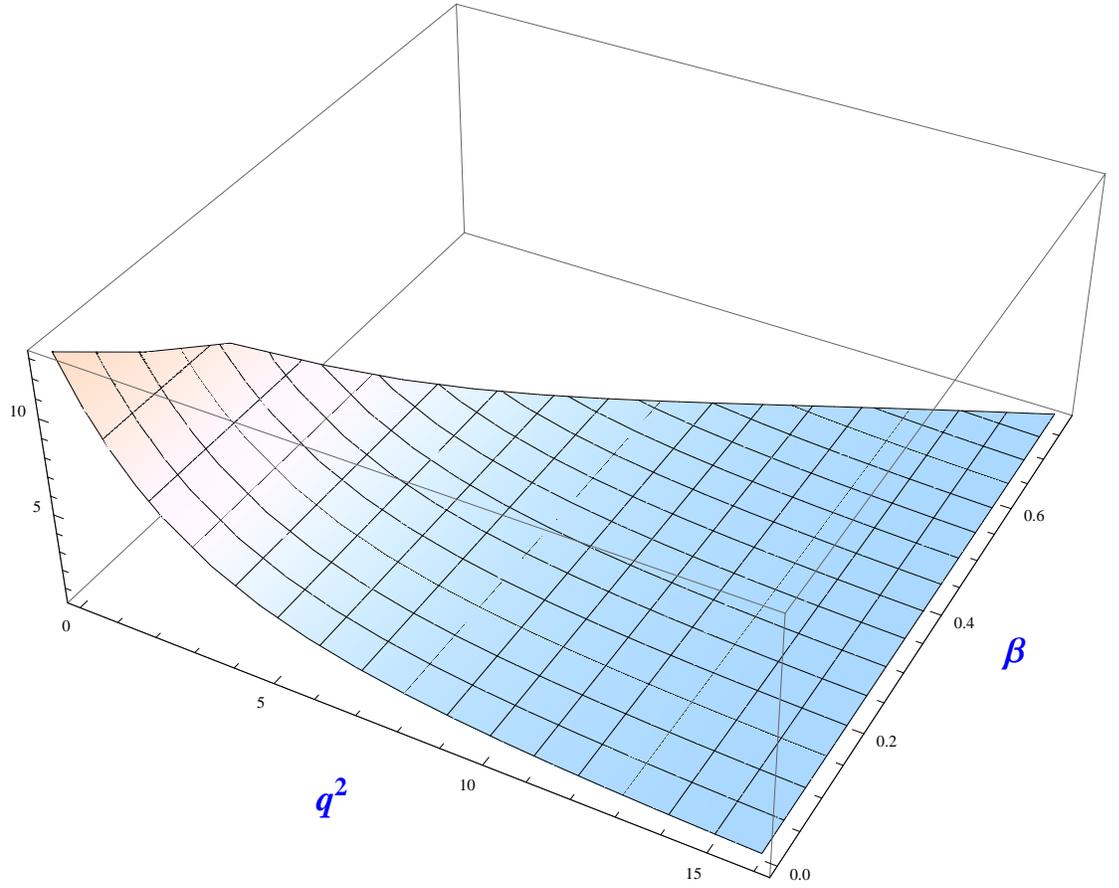}
\label{F1}
\end{center}
\end{figure}

\begin{figure}
\begin{center}
\caption{Distribution  $d\sigma^{2+}/dq^2d\beta$ in $10^{-10}r_0^4$,
$r_0=1/m\alpha$ for $\beta=0$. Solid line is for $\omega=800$\,eV,
dashed line is for $\omega=1$ keV. The recoil momentum $q$ is in atomic units.}
\epsfxsize=20cm\epsfbox{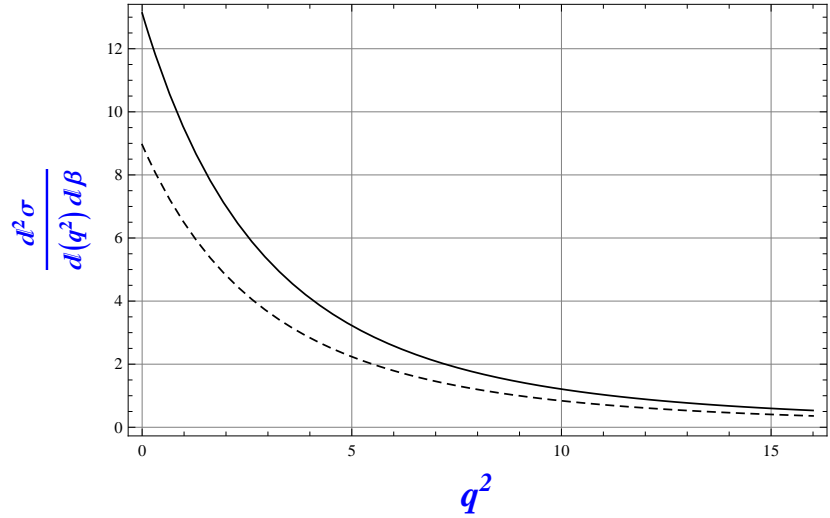}
\label{F2}
\end{center}
\end{figure}

\begin{figure}
\begin{center}
\caption{Distribution  $d\sigma^{2+}/dq^2$ in
$10^{-10}r_0^4$, $r_0=1/m\alpha$. Notations are the same as in Fig.\,2}
\epsfxsize=20cm\epsfbox{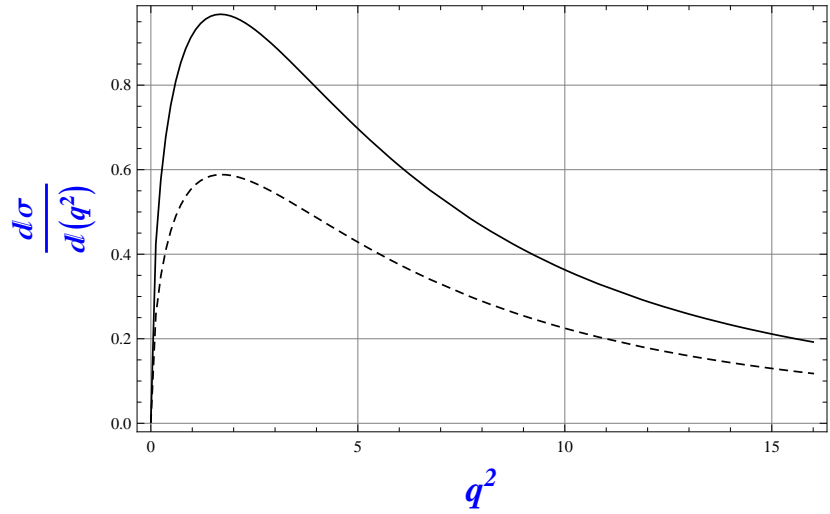}
\label{F3}
\end{center}
\end{figure}

\begin{figure}
\begin{center}
\caption{Distribution  $d\sigma^{2+}/dtd\beta$ in $barns$
for $\omega=800$\,eV.}
\epsfxsize=20cm\epsfbox{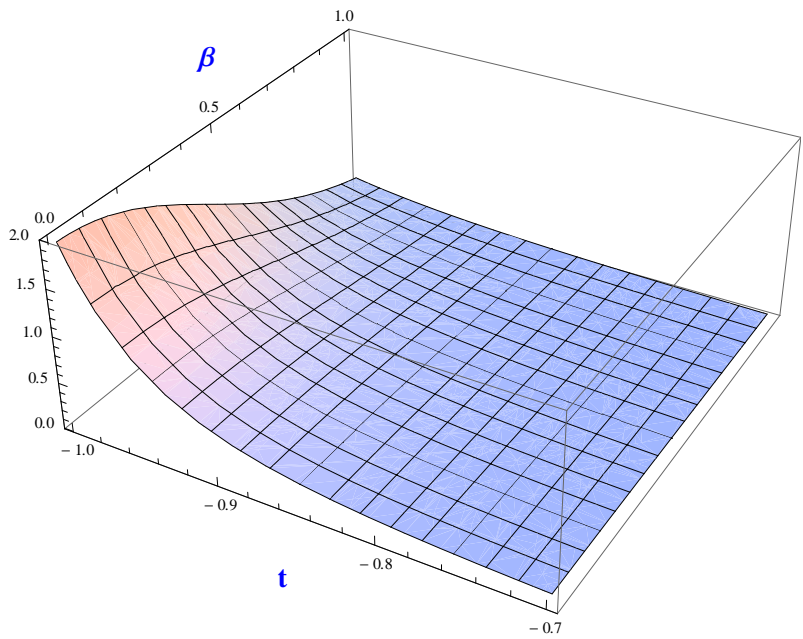}
\label{F4}
\end{center}
\end{figure}

\begin{figure}
\begin{center}
\caption{Distribution  $d\sigma^{2+}/dtd\beta$ in $barns$
for $\beta=0$. Notations are the same as in Fig.\,2.}
\epsfxsize=20cm\epsfbox{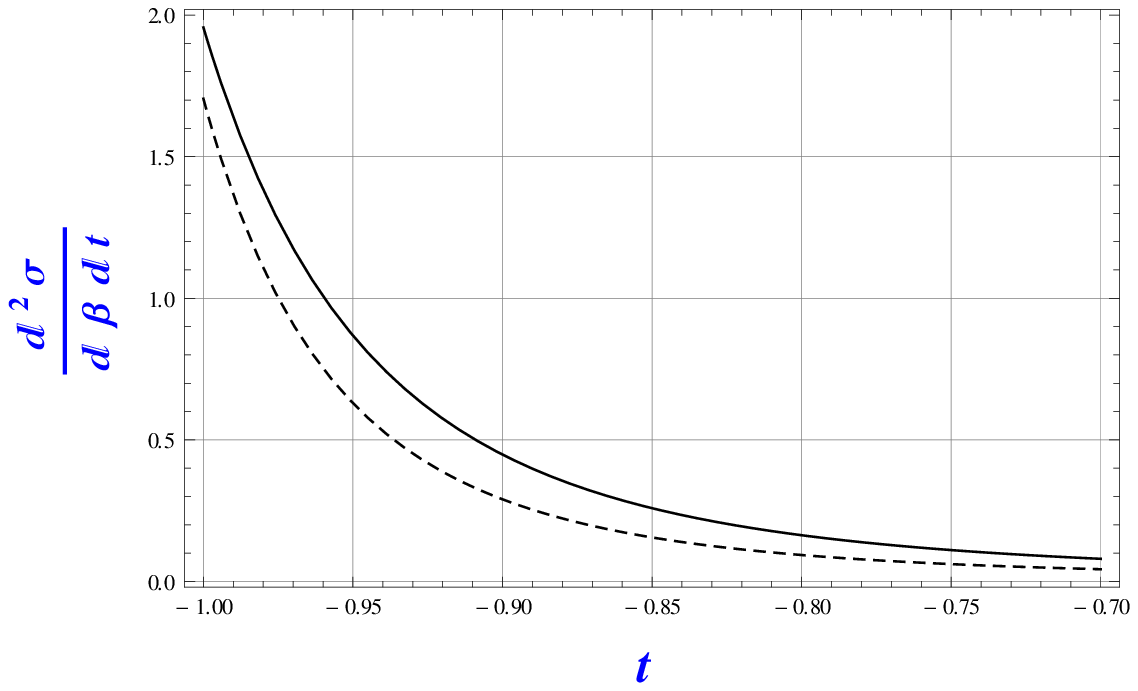}
\label{F5}
\end{center}
\end{figure}

\begin{figure}
\begin{center}
\caption{Distribution  $d\sigma^{2+}/dt$ in $barns$
for $\beta=0$. Notations are the same as in Fig.\,2.}
\epsfxsize=20cm\epsfbox{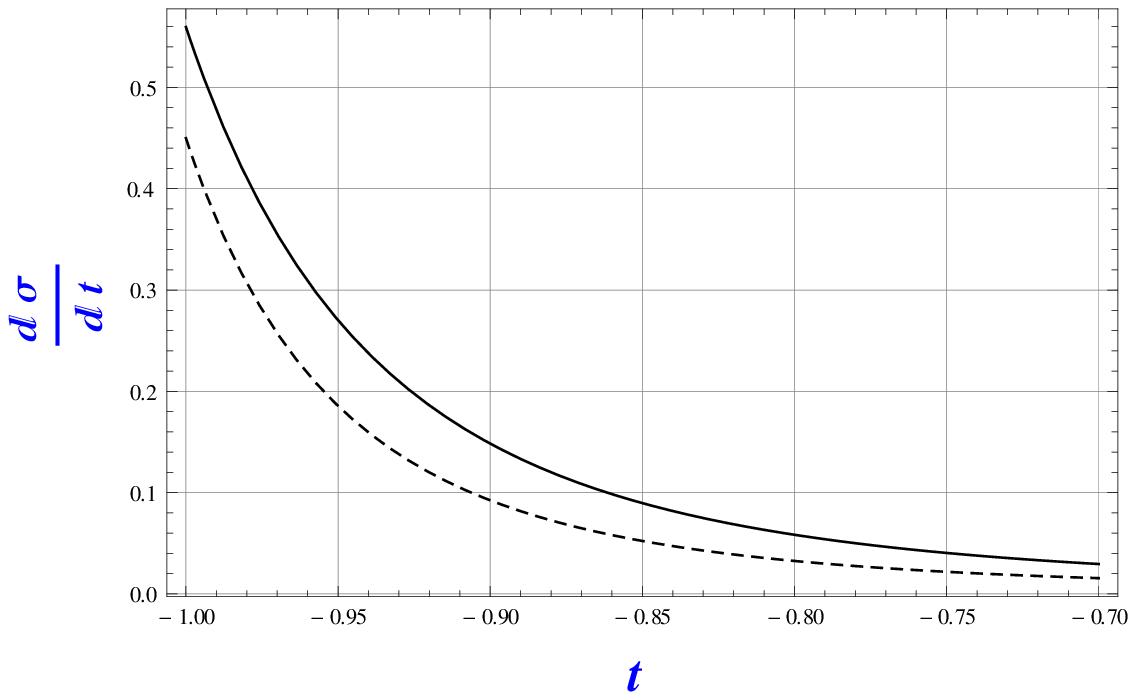}
\label{F6}
\end{center}
\end{figure}


\begin{thebibliography}{99}


\bibitem{7a} M. S. Sch\"{o}ffler \textit{et al.}, ICPEAC
2011(http://www.qub.ac.uk/icpeac 2011/).

\bibitem{7b}  Th. Weber \textit{et al.}, Bull. Amer. Phys. Soc. {\bf 56}, n.5,
144(2011).

\bibitem{7} M. S. Sch\"{o}ffler \textit{et al.} arXiv: 1207.7181 [physics.atom-ph] (2012).

\bibitem{1}  M. Ya. Amusia, E. G. Drukarev, V. G. Gorshkov, and M. P.
Kazachkov, J.Phys. B {\bf 8}, 1248 (1975).

\bibitem{2} Z.J.Teng, R. Shakeshaft,  Phys. Rev. A {\bf 49},3597 (1994).

\bibitem{3} E. G. Drukarev, Phys. Rev. A {\bf 52}, 3910 (1995).

\bibitem{4} E. G. Drukarev, N. B. Avdonina and R. H. Pratt, J. Phys.B  {\bf
34}, 1 (2001).

\bibitem{5}  J. A. Ludlow, J. Colgan, T.G. Lee, M. S. Pindzola, and F. Robicheaux,, J. Phys. B. {\bf 42} 225204 (2009).

\bibitem{6} A. G. Galstyan, O. Chuluunbaatar, Yu. V. Popov, and B. Piraux, Phys. Rev. A {\bf 85}, 023418 (2012).

\bibitem{8} T. Suric, E. G. Drukarev, and R. H. Pratt, Phys. Rev. A {\bf 67},
022709 (2003).

\bibitem{9} T. Kato, Commun. Pure Appl.Math. {\bf 10}, 151 (1957).

\bibitem{10} E. Z. Liverts, N. Barnea, Comp. Phys. Com. {\bf 182}, 1790
(2011).

\bibitem{11} E. Z. Liverts,  M. Ya. Amusia, R. Krivec, and V.B. Mandelzweig,
Phys. Rev. A {\bf 73}, 012514 (2006).

\bibitem{12} M. Ya. Amusia, E. G. Drukarev, and E. Z. Liverts, JETP Letters, {\bf 103}, 690 (2012).

\bibitem{13}  A. Nordsieck, Phys. Rev.{\bf 93}, 785 (1954).

\bibitem{14}  M. Ya. Amusia, E. G. Drukarev, V.B. Mandelzweig, Phys. Scr. {\bf
72}, C22 (2005).

\bibitem {15} E. G. Drukarev, Physics-Uspekhi, {\bf 50}, 835 (2007).




\end{thebibliography}
\end{document}